\documentclass[prl,aps,twocolumn,superscriptaddress,floatfix,showpacs]{revtex4}
\usepackage{graphicx,rotating,subfigure,amsmath,amsfonts,amssymb,delarray}
\usepackage[T1]{fontenc}

\newcommand{\e}{\text{e}}

\def\l{\left}
\def\r{\right}
\def\12{\frac{1}{2}}
\def\nn{\nonumber}
\newcommand{\be}{\begin{equation}}
\newcommand{\ee}{\end{equation}}
\newcommand{\bea}{\begin{eqnarray}}
\newcommand{\eea}{\end{eqnarray}}
\newcommand{\ket}{\rangle}
\newcommand{\bra}{\langle}
\newcommand{\J}{\mathcal{J}}
\renewcommand\Re{\operatorname{Re}}

\begin{document}
\bibliographystyle{apsrev}

\title{Diffusion and ballistic transport in one-dimensional quantum systems}

\author{J. Sirker} 
\affiliation{Department of Physics and Research Center OPTIMAS, University of Kaiserslautern, D-67663
  Kaiserslautern, Germany}

\author{R. G. Pereira}
\affiliation{Kavli Institute for Theoretical Physics, University of
  California, Santa Barbara, CA 93106, USA}

\author{I. Affleck}
\affiliation{Department of Physics and Astronomy, University of British
  Columbia, Vancouver, BC, Canada V6T1Z1}

\date{\today}

\begin{abstract}
  It has been conjectured that transport in integrable one-dimensional (1D)
  systems is necessarily ballistic. The large diffusive response seen
  experimentally in nearly ideal realizations of the $S=1/2$ 1D Heisenberg
  model is therefore puzzling and has not been explained so far. Here, we show
  that, contrary to common belief, diffusion is universally present in
  interacting 1D systems subject to a periodic lattice potential. We present a
  parameter-free formula for the spin-lattice relaxation rate which is in
  excellent agreement with experiment. Furthermore, we calculate the current
  decay directly in the thermodynamic limit using a time-dependent density
  matrix renormalization group algorithm and show that an anomalously large
  time scale exists even at high temperatures.
\end{abstract}

\pacs{72.10.-d, 05.60.Gg, 05.10.Cc, 75.40.Gb}

\maketitle


For a generic system of interacting particles at sufficiently high
temperatures, transport is expected to be scattering limited.  In $d$
spatial dimensions, the signature of diffusive motion is the
characteristic long-time decay of the autocorrelation function
$\langle n_\mathbf{r}(t) n_\mathbf{r}(0) \rangle\sim t^{-d/2}$. Here,
$n_\mathbf{r}$ represents the density of a globally conserved quantity
$\sum_\mathbf{r}n_\mathbf{r}$. In very clean systems, however,
transport can be a subtle issue because constants of motion may slow
down the current decay or even prevent currents from decaying
completely. 
An important role in our understanding of strongly correlated
electrons is played by {\it integrable} quantum models. Since these
models possess an infinite number of {\it local} conserved quantities,
one might expect ideal ({\it ballistic}) transport to be the rule
rather than the exception \cite{CastellaZotos}. Whether or not
diffusive behavior is possible at all in such systems is indeed an
intensely studied
\cite{CastellaZotos,RoschAndrei,FujimotoKawakami,Zotos,AlvarezGros,ZotosPrelovsek,NarozhnyMillis,HeidrichMeisner2,JungRoschDrude,KluemperJPSJ,FabriciusMcCoy,SirkerDiff,GiamarchiMillis}
but still open question. Experimentally, the question if spin
diffusion holds in Heisenberg chains has been investigated for decades
\cite{SteinerVillain,BoucherBakheit,ThurberHunt,PrattBlundell}.

\begin{figure}[t!]
\begin{center}
\includegraphics*[width=1.0\columnwidth]{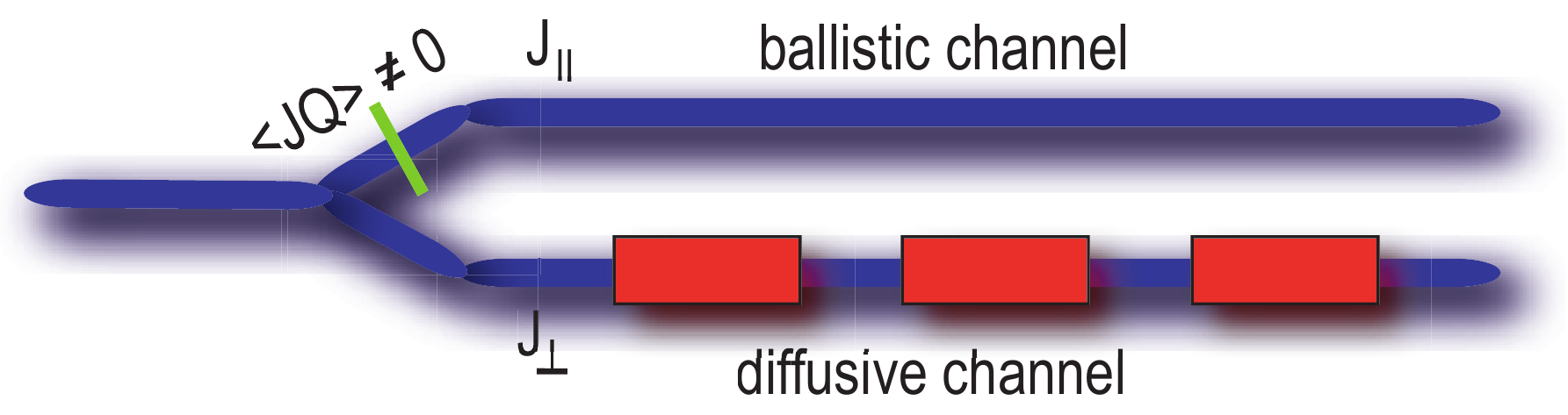}
\end{center}
\caption{In a diffusive channel, the conductivity is limited by the dominant
  of the various scattering processes pictured as a serial arrangement of
  resistors. If part of the current is, however, protected by a conservation
  law, a parallel ballistic channel for charge transport is opened.}
\label{illu}
\end{figure}

In the thermodynamic limit, ballistic transport can be defined from
the condition that the current-current correlation function
$\langle\J(t)\J(0)\rangle$, where $\J$ is the spatial integral of the
current density operator and the brackets denote thermal average, does
not decay to zero at large times.  This happens, for example, in a
free electron gas, where $\J$ is proportional to the momentum operator
and therefore conserved in a translationally invariant system
\cite{GiamarchiMillis}. The dc conductivity is then infinite. 
Next, we consider the case where the current operator itself is not
conserved but a conserved quantity $Q$ exists which has finite overlap with
$\J$.
We can then write $\J=\J_\parallel + \J_\perp$, with $\J_\parallel = (\langle
\J Q\rangle/\langle Q^2\rangle)Q$ being the part which cannot decay
\cite{JungRoschDrude}, 
leading to parallel diffusive and ballistic channels as indicated in
Fig.~\ref{illu}. 
This idea can be generalized to a set of orthogonal conserved
quantities $Q_n$, $\langle Q_nQ_m\rangle = \langle Q_n^2\rangle \delta_{n,m}$,
and leads to Mazur's inequality \cite{Mazur,ZotosPrelovsek}
\begin{equation}
\label{Mazur}
D = \frac{1}{2LT}\lim_{t\to\infty} \langle \J(t)\J(0)\rangle
\geq\frac{1}{2LT}\sum_n\frac{\langle 
\J Q_n\rangle^2}{\langle Q_n^2\rangle} \, .
\end{equation}
Here, $L$ is the system size and $T$ the temperature.  
The {\it Drude weight} $D$ measures the weight of the delta-function peak in
the real part of the optical conductivity at zero frequency,
$\sigma'(\omega)=2\pi D\delta(\omega)+\sigma_{reg}(\omega)$. In principle,
both $D$ and $\sigma_{reg}(\omega=0)$ can be nonzero \cite{RoschAndrei}. Weak
breaking of the conservation laws renders the conductivity finite, but in this
case the 
projection of the current onto the longest lived $Q_n$ sets a lower bound for
the conductivity \cite{JungRoschDrude}.

It is important to note that the rhs of Eq.~(\ref{Mazur}) can
vanish even if integrability allows us to construct an infinite set of
conserved quantitities. In the following, we consider the integrable
model of {\it spinless} fermions (XXZ model)
\begin{equation}
\label{HXXZ}
H = J \sum_{l=1}^N \big[-\frac{1}{2}\left(c^\dagger_lc^{\phantom\dagger}_{l+1}+ h.c.\right)+\Delta
(n_{l}-\frac{1}{2})(n_{l+1}-{\frac{1}{2}})\big].
\end{equation}
Here $N$ is the number of sites, $J$ the hopping amplitude,  $c_l$ annihilates
a fermion at site $l$, and $\Delta$ is the interaction strength.
This model is equivalent to the anisotropic spin-1/2 chain and is
exactly solvable by Bethe ansatz (BA) \cite{GiamarchiBook}.  At half-filling, $\bra
n_l\ket=1/2$, the excitation spectrum is gapless for $|\Delta|\leq 1$ and
gapped for $|\Delta|>1$. 
The current operator is  $\J=\sum_l j_l$, with $j_l = -iJ(c_l^\dagger c^{\phantom\dagger}_{l+1} - c_{l+1}^\dagger
c^{\phantom\dagger}_l)/2$ as follows from a discretized continuity equation.

At zero temperature, the Drude weight can be calculated by BA 
\cite{ShastrySutherland} and is found to be finite in the gapless and zero in
the gapped regime. Mazur's inequality can be used to show that $D(T)\neq 0$
away from half-filling at arbitrary temperatures \cite{ZotosPrelovsek}.
Remarkably, at half-filling the Mazur bound for the Drude weight obtained from
{\it all local} conserved quantities vanishes identically due to particle-hole
symmetry. Since this is only a lower bound, it does not imply that $D$ itself
vanishes. However, one can argue
\cite{DamleSachdev} that in the gapped regime $D$ should remain zero
at finite temperatures. The main open question is whether the Drude
weight is finite at finite temperatures in the half-filled gapless
case. Since Eq.~(\ref{Mazur}) is actually an equality if all
conserved quantities are included \cite{Suzuki}, a nonzero $D$ at
half-filling requires the existence of a \emph{nonlocal} conserved quantity which has finite overlap with the current operator
\cite{JungRoschDrude}. $D(T>0)\neq 0$ at half-filling has been found
in two independent BA calculations \cite{Zotos,KluemperJPSJ}. However,
these results disagree and they both violate exact relations for
$D(T)$ at high temperatures \cite{KluemperJPSJ}. Further evidence for
$D(T>0)\neq 0$ stems from exact diagonalization (ED)
\cite{NarozhnyMillis,HeidrichMeisner2,JungRoschDrude} and Quantum
Monte Carlo (QMC) \cite{AlvarezGros,HeidarianSorella}. We will discuss
these numerical works in relation to our own results at the end of
this letter.

Evidence for diffusion in  the spin-spin autocorrelation
function at high temperatures 
has been sought via ED \cite{FabriciusMcCoy}, QMC \cite{StarykhSandvik} and density
matrix renormalization group (DMRG) 
\cite{SirkerKluemperDTMRG,SirkerDiff}. 
The results at infinite temperature seemed
consistent with an algebraic decay $\langle S^z_l(t)S^z_l(0)\rangle \sim
t^{-\alpha}$ with exponent $\alpha$ close to $1/2$ as expected for $d=1$. At
low temperatures, the diffusive contribution 
was practically undetectable \cite{SirkerDiff}.  Meanwhile, nuclear magnetic
resonance (NMR) \cite{ThurberHunt} and muon spin relaxation
\cite{PrattBlundell} experiments even found evidence for {\it low-temperature
  diffusive behavior} in two completely different $S=1/2$ Heisenberg chain
compounds, but have so far remained unexplained.

In the NMR experiment on the spin chain compound Sr$_2$CuO$_3$, spin diffusion
is observed as a characteristic magnetic field dependence of the spin lattice
relaxation rate, $1/T_1\sim 1/\sqrt{h}$ \cite{ThurberHunt}. Here, only
excitations with momentum $q\sim 0$, relevant for the studied transport
properties, contribute. Clearly, Sr$_2$CuO$_3$ is not exactly an integrable
system. However, the behavior is expected to be different depending on
whether the diffusion constant is determined by intrinsic umklapp scattering
within the integrable model or by integrability-breaking perturbations. The
spin excitations propagating in a given channel only contribute to the
diffusive response at frequencies which are small compared to the relaxation
rate in that channel. If the Drude weight of the XXZ model is large in the
regime $h\ll T\ll J$, then we expect a large fraction of the excitations in
Sr$_2$CuO$_3$ to propagate in a quasi-ballistic channel with a very small
relaxation rate. The diffusive response should therefore be suppressed
compared to the case where the integrable model has a dominant diffusive
channel.

\begin{figure}[t!]
\includegraphics*[width=\columnwidth]{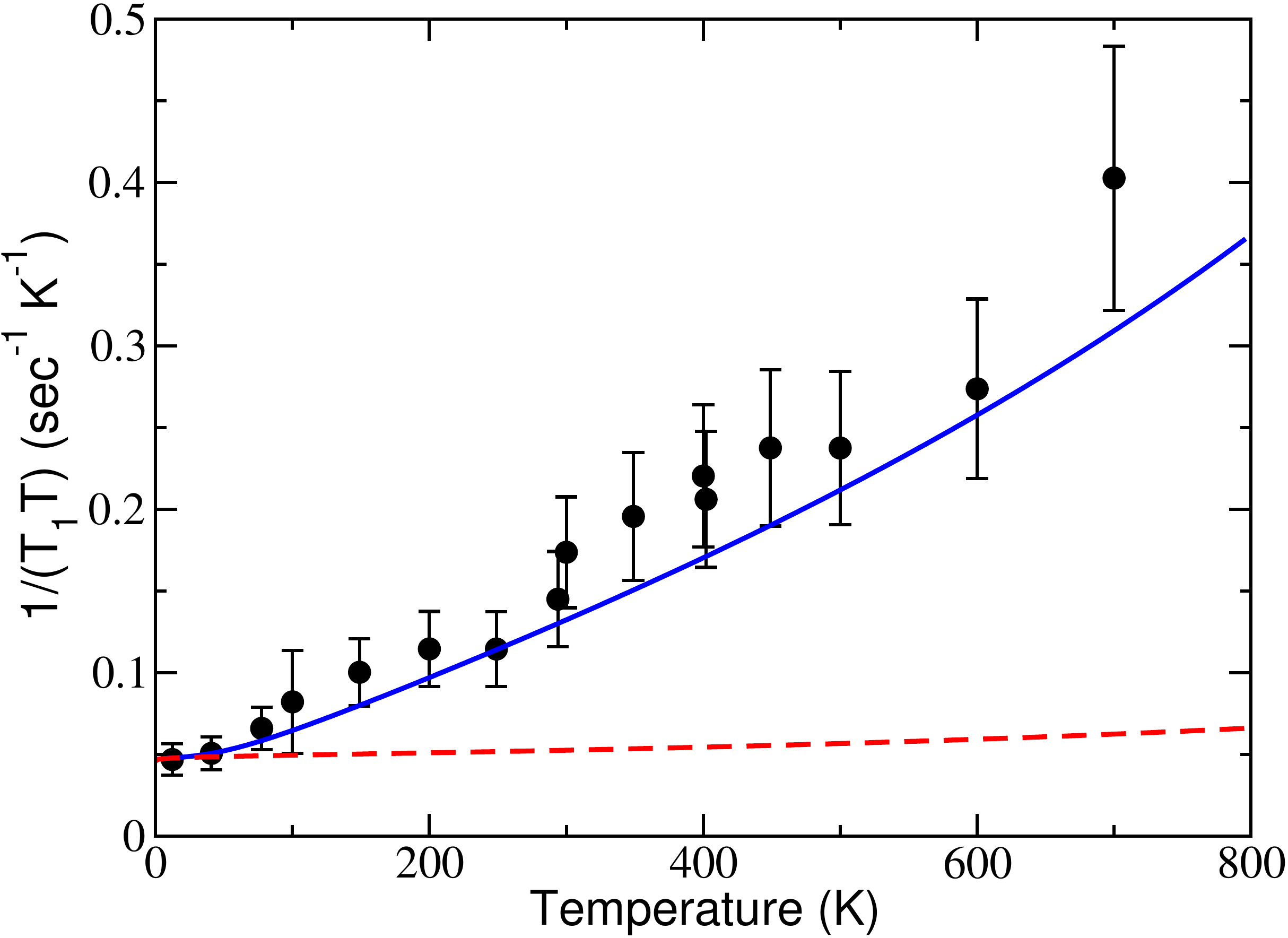}
\caption{Experimental data for $1/T_1$ of the spin chain
  compound Sr$_2$CuO$_3$ at $h=9$ T taken from Ref.~[\onlinecite{ThurberHunt}] (dots)
  compared to our theory (blue solid line). Without diffusion, $\gamma=0$,
  $1/(T_1T)$ would be almost constant (red dashed line).} 
\label{Fig_T1}
\end{figure}

We now calculate $1/T_1$ by a standard field theory approach based on
the Luttinger model \cite{GiamarchiBook} assuming that there is no
unknown nonlocal conservation law that has a finite overlap with
$\J$. For $T\gg\omega_e$ and $\Delta=1$ we have
\begin{equation}
\label{T1}
\frac{1}{T_1} \approx -\frac{2T}{\omega_e}\int\frac{dq}{2\pi} |A(q)|^2 \,\chi''_{\rm
  ret}(q,\omega_e) \; .
\end{equation}
Here $\chi''_{\rm ret}(q,\omega)$ is the imaginary part of the {\it
  longitudinal} retarded spin-spin correlation function $\chi_{\rm
  ret}(q,\omega)$ and $\omega_e=\mu_Bh$. $1/T_1$ is determined by the
{\it transverse} spin Green's function at the {\it nuclear} resonance
frequency, $\omega_N\approx 0$. By including the Zeeman term in the
time evolution of the transverse spin operator but ignoring its
negligible effects on the Boltzmann weights and using the resulting
$SU(2)$ symmetry we express $1/T_1$ in terms of the {\it longitudinal}
Green's function at the {\it electron} resonance frequency $\omega_e$
in (\ref{T1}). For the in-chain oxygen site in Sr$_2$CuO$_3$, we have
$A(q)=A\cos(q/2)$ with
$|A|^2=k_B(g\gamma_N\hbar)^2[(2C^b)^2+(2C^c)^2]/(2\hbar\pi^3
k_B^2J^2)$ where $k_B$ is the Boltzmann constant, $C^{b,c}$ are the
dimensionless components of the hyperfine coupling tensor,
$g\gamma_N\hbar = 4.74\times 10^{-9}$ eV and $J$ is the exchange
coupling measured in Kelvin. To obtain the curve shown in
Fig.~\ref{Fig_T1}, we used $J=2000$ K and $2C^b=105$, and $2C^c=54$
\cite{ThurberHunt}. For small momentum $q$ we find
\begin{equation}
\label{chiret}
\chi_{\rm ret}(q,\omega) = \frac{vKq^2}{2\pi}\frac{1}{\omega^2-v^2q^2-\Pi_{\rm
    ret}(q,\omega)} \; .
\end{equation}
Here $K$ is the Luttinger parameter and $v$ the spin velocity. For the pure
Luttinger model, $\Pi_{\rm ret}(q,\omega)\equiv 0$, leading to $1/T_1\sim
T$ in the limit $T\to0$ \cite{SachdevNMR}. In second order in the umklapp
scattering and first order in band curvature the self-energy has the
form
\begin{equation}
\label{selfE}
\Pi_{\rm ret}(q,\omega) \approx -2i\gamma\omega -b\omega^2+cv^2q^2 \; .
\end{equation}
For the experimentally relevant isotropic case ($\Delta=1$), $K\approx
1+g/2$ and $v=J\pi/2$. For the decay rate $\gamma(T)$ and the
parameters $b$ and $c$ we find in this case
\begin{eqnarray}
\label{selfE_parameters_iso}
2\gamma &=& \pi g^2T ,\quad c = \frac{g^2}{4}-\frac{3g^3}{32} -\frac{\sqrt{3}}{\pi}T^2, \\
b &=&
\underbrace{\frac{g^2}{4}-\frac{g^3}{32}\l(3-\frac{8\pi^2}{3}\r)}_{b_2}+\underbrace{\frac{\sqrt{3}}{\pi}T^2}_{b_1}
. \nn
\end{eqnarray}
Following Lukyanov \cite{Lukyanov}, the running coupling constant $g(T)$ is determined by the equation 
\begin{equation}
\label{coup_g}
\frac{1}{g}+\frac{\ln g}{2} = \ln\l[\sqrt{\frac{\pi}{2}}\frac{\e^{1/4+\tilde{\gamma}}}{T}\r],
\end{equation}
where $\tilde{\gamma}$ is the Euler constant. Similarly, the parameters
$\gamma,\, b$ and $c$ can be determined for the anisotropic case
$0<\Delta<1$ (see EPAPS document No.). Importantly, we always find a finite
decay rate implying spin diffusion in the sense that $\langle
n_l(t)n_l(0)\rangle \sim T\sqrt{\gamma/t}$ at large times.
At high temperatures such that $\gamma\gg h$ but still
$T\ll J$, we find $1/T_1\sim T\sqrt{\gamma/h}$ with $\gamma\sim T/\ln^2(J/T)$
for the isotropic Heisenberg model. A comparison of the essentially
parameter-free calculated temperature dependence with experiment is shown in
Fig.~\ref{Fig_T1}.
The good agreement indicates that a large diffusive response is present in the
integrable Heisenberg model near half-filling. Furthermore, this result shows that
umklapp scattering is a ``dangerously irrelevant'' perturbation of the
Luttinger model \cite{SachdevNMR}, completely changing the behavior of $1/T_1$
in the regime $h \ll T$ from a constant to a square-root divergence
$1/\sqrt{h}$, as seen in experiment.

Our field theory calculation assumed $D(T>0)=0$. The optical conductivity
$\sigma(q,\omega) =i\omega\chi_{\rm ret}(q,\omega)/q^2$ can be obtained from
Eq.~(\ref{chiret}) and we find that 
\begin{equation}
\label{sigma_selfE}
\sigma'(\omega) = \frac{vK}{2\pi}\frac{2\gamma}{[(1+b)\omega]^2+(2\gamma)^2}
\end{equation}
is a Lorentzian with width set by $\gamma$.  If conservation laws
protecting the Drude weight are present, they can be naturally incorporated
using the memory matrix formalism \cite{RoschAndrei}. For a single conservation
law $[Q,H]=0$, this formalism yields
\begin{eqnarray}
\label{condMM}
\sigma'(\omega) &=& \frac{Kv}{2\pi(1+y)}\bigg[ \pi y(1-b_1)\delta(\omega)\\
&+&\frac{2\gamma'}{[(1+b_1+b_2^\prime)\omega]^2+(2\gamma')^2}\bigg]\nonumber  ,
\end{eqnarray}
where the parameter $y\equiv\langle \J Q\rangle^2/(\langle \J^2\rangle \langle
Q^2\rangle -\langle \J Q\rangle^2 )$ measures the overlap of $\J$ with the
conserved quantity, and $\gamma'=(1+y)\gamma$ and $b_2^\prime=(1+y)b_2$. Here
$b_1$ and $b_2$ in the isotropic case are the parameters defined in
Eq.~(\ref{selfE_parameters_iso}). Note that for $y=0$, (\ref{condMM}) reduces
to the optical conductivity (\ref{sigma_selfE}) obtained in the self-energy
approach.  According to Eq.~(\ref{condMM}), $\sigma'(\omega)$ has a ballistic
and a regular (diffusive) part, with the weight in each part being controlled
by $y$. Away from half-filling (finite magnetic field in the spin chain), a
lower bound for $y$ is provided by the overlap with the conserved energy
current operator $Q=\J_E$ \cite{CastellaZotos}. In this case, $y\sim (h/T)^2$.
In the half-filled case a possible unknown nonlocal conservation law would
mean that spectral weight is shifted from the Lorentzian into a ballistic part
that does not contribute to the temperature dependence of $1/(T_1T)$ (see
dashed line in Fig.~\ref{Fig_T1}). That the experimental points in
Fig.~\ref{Fig_T1} are actually mostly \emph{above} the theoretical prediction
suggests that $D$ is rather small near half-filling.  

\begin{figure*}[ht]
\includegraphics*[width=0.99\columnwidth]{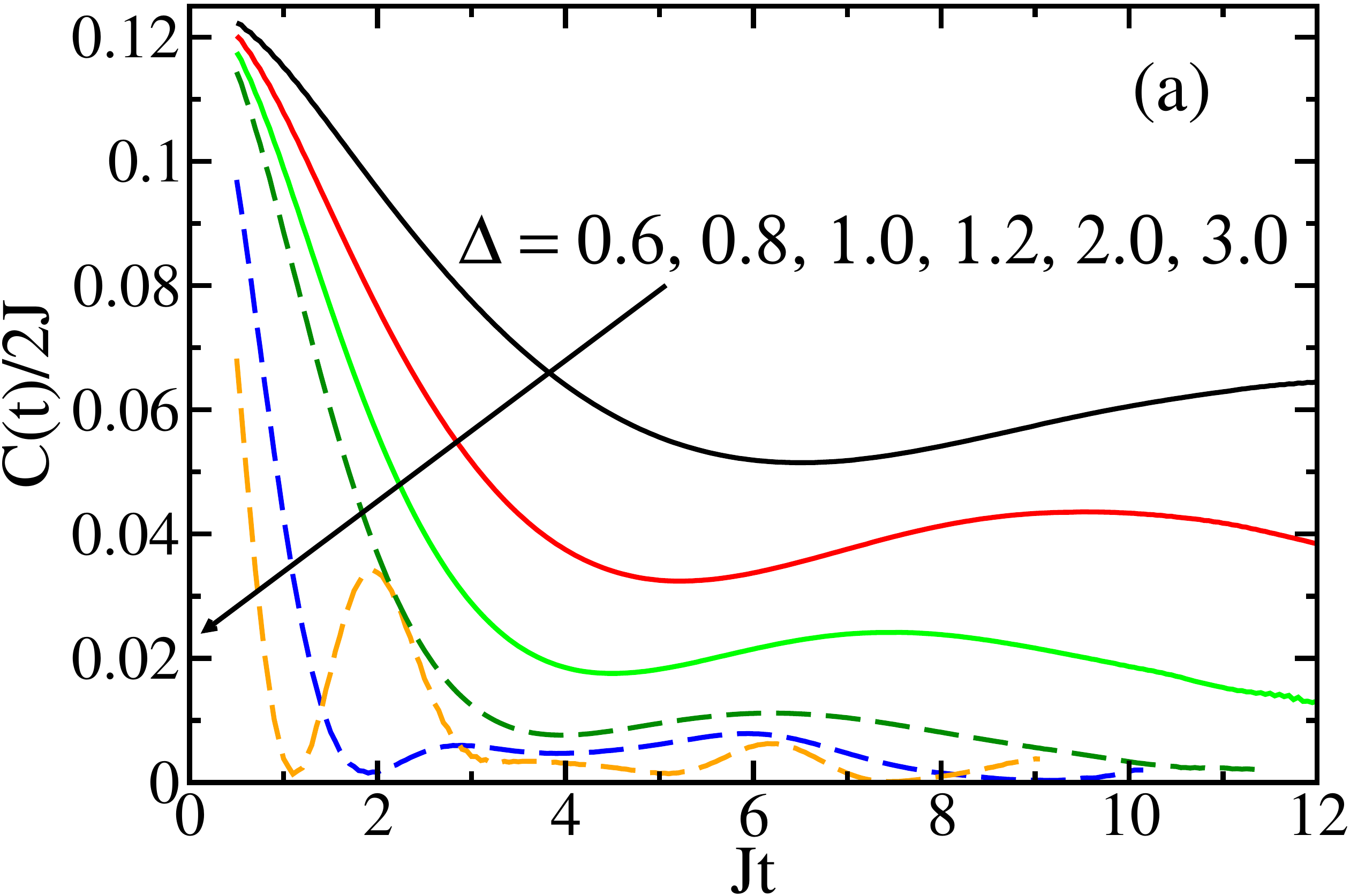}
\includegraphics*[width=0.99\columnwidth]{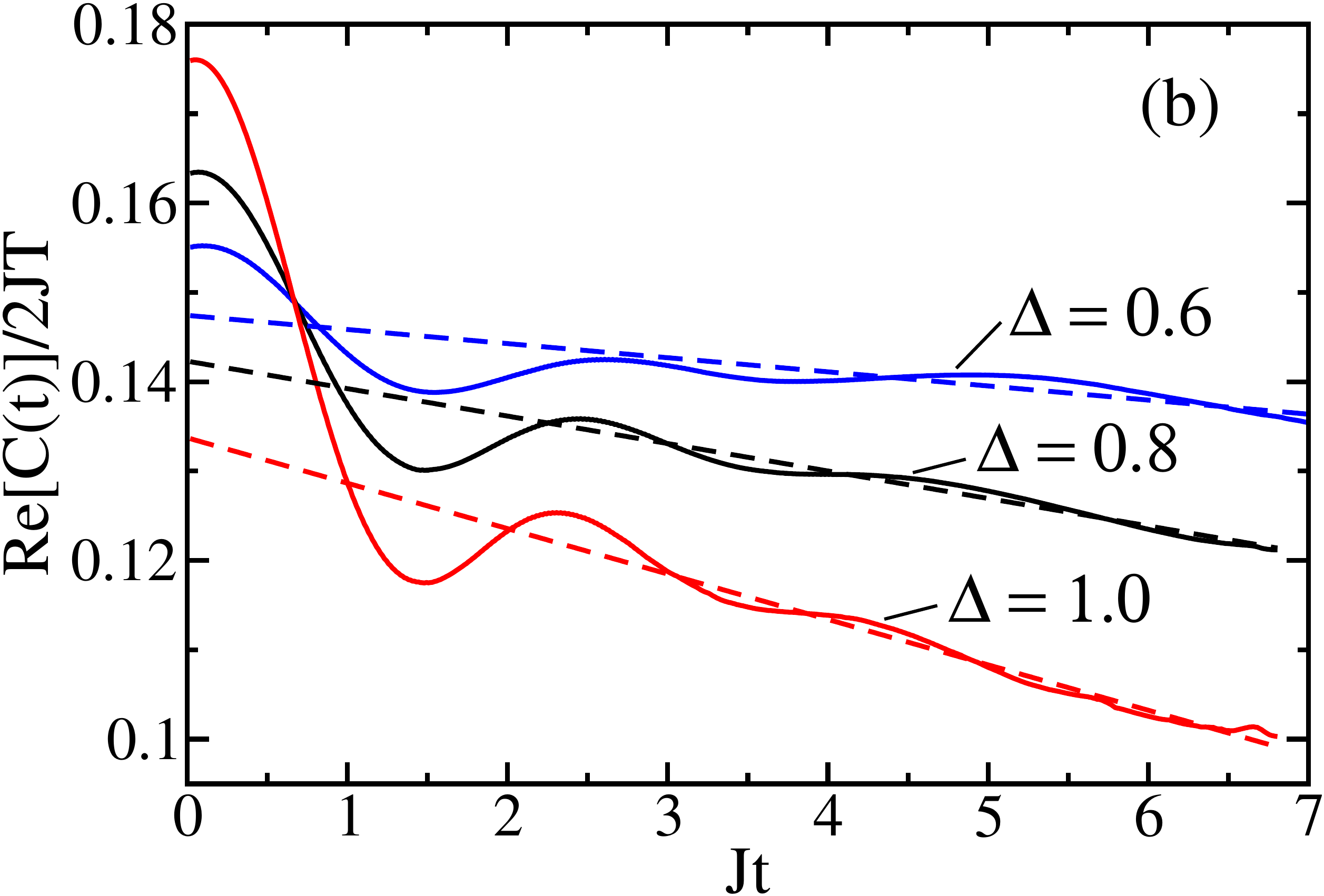}
\caption{(a) $C(t)=\langle \J(t)\J(0)\rangle/L$ at $T=\infty$ for various
  $\Delta$ as indicated on the plot. The solid (dashed) lines correspond to
  $\Delta$ in the critical (gapped) regime, respectively. (b) $\Re [C(t)]/2JT$
  at $T=0.2 J$ for $\Delta=0.6$ (blue solid line), $\Delta=0.8$ (black solid
  line), and $\Delta=1.0$ (red solid line). The dashed lines are linear fits
  $\Re [C(t)]/2JT=A-BJt$ for $Jt\in [3.5,7]$.}
\label{Fig_TMRG}
\end{figure*}

In order to clarify the contradiction with previous studies that supported a
large Drude weight at half-filling \cite{HeidrichMeisner2,NarozhnyMillis}, we
used a DMRG algorithm \cite{SirkerKluemperDTMRG,SirkerDiff} to calculate the
current-current correlation $C(t)\equiv \langle \J(t)\J\rangle/L$ directly in
the thermodynamic limit. According to Eq.~(\ref{Mazur}), this correlation
function asymptotically yields $D$. Remarkably, the results in
Fig.~\ref{Fig_TMRG}(a) show that $C(t)$ is nonmonotonic and does not converge
to an asymptotic value for times up to $Jt=11$ even for infinite temperature.
This is true within the critical as well as the gapped regime. Note
that the time scales reached in our DMRG calculations are about a
factor of $2$ larger than what can be achieved by ED where only times
$vt<N/2$ are accessible. We conclude that a large time
scale 
persists at $T=\infty$ posing a serious challenge for ED studies.

While previous QMC results \cite{AlvarezGros} are unable to resolve
the small decay rate, $\gamma(T)\ll T$, very recent ones
\cite{GrossjohannBrenig} seem to strongly support our expression for
$\gamma(T)$ in Eq.~(\ref{selfE_parameters_iso})
\cite{GrossjohannBrenig}. Further evidence that $\gamma(T)$ is nonzero
for $T\ll J$ is provided by Fig.~\ref{Fig_TMRG}(b) showing $\Re
[C(t)]/2JT$ at $T=0.2J$. The result in Eq.~(\ref{condMM}) predicts for
the decay of the current-current correlation function for $t\gg (2\pi
T)^{-1}$ and neglecting the small imaginary part (suppressed by a
factor $\gamma/T$): \be
C(t)\sim\frac{vKT}{2\pi(1+y)}\l(y(1-b_1)+\frac{e^{-2\gamma^\prime
    t}}{1+b_1+b_2^\prime}\r).  \ee At intermediate times $(2\pi
T)^{-1}\ll t\ll 1/\gamma'$ we obtain a linear decay independent of $y$
if $b_1,b_2^\prime\ll 1$ \be
\label{lin_decay}
C(t)\approx KvT(1-2\gamma t)/[2\pi(1+b)] \; .
\ee
A linear fit in this regime yields values which are
consistent with our theory (see table \ref{tab1}).
\begin{table}
\begin{ruledtabular}
\begin{tabular}{ccccc}
$\Delta$ & $A$ & $vK/4\pi(1+b)$ & $B/2A$ & $\gamma$\\
\hline\\*[-0.15cm]
0.6 &0.147  & 0.147 &0.0054  & 0.0052\\
0.8 &0.142  & 0.140 &0.0109  & 0.0116\\
1 &0.134  & 0.135 &0.0190  & 0.0297
\end{tabular}
\end{ruledtabular}
\caption{Parameters obtained by fitting $\Re [C(t)]/T$ in Fig.~\ref{Fig_TMRG}(b) to $\Re [C(t)]/T=A-Bt$.
  According to Eq.~(\ref{lin_decay}), we expect $A=vK/\pi(1+b)$ and
  $B/A=2\gamma$ with parameters $\gamma$ and $b$ as given in Eq.~(\ref{selfE_parameters_iso}) for
  $\Delta=1$ and in the EPAPS document No. for the anisotropic case,
  respectively.}
\label{tab1}
\end{table}
We also note that the values of $C(t)/2JT$ for $Jt\approx 6$ are already
smaller than the Drude weight found in [\onlinecite{KluemperJPSJ}] by BA. 

To summarize, we have shown that in integrable 1D systems diffusion can
coexist with ballistic transport, in the sense illustrated in Fig.~\ref{illu}.
This is the scenario for the XXZ model away from half-filling. For the
half-filled case, however, we have argued that the large diffusive response
measured experimentally in spin chains and seen in our numerical calculations
suggests that, contrary to common belief, the low-temperature Drude weight is
either zero or surprisingly small for $\Delta$ near 1. 

\acknowledgments The authors thank A.~Alvarez and C.~Gros for sending
us their quantum Monte Carlo data and acknowledge valuable discussions
with T.~Imai, A.~Kl\"umper and A.~Rosch. This research was supported
by NSERC (J.S., R.G.P., I.A.), CIfAR (I.A.), the NSF under Grant No.
PHY05-51164 (R.G.P.), and the Research Center OPTIMAS (J.S.).


\end{document}